\documentclass{elsart}
\usepackage{epsfig}

\def\hc#1{\hbox to\hsize{\hss #1\hss}}

\journal{{\tt Astroparticle Physics}}

\let\mal=\times

\def\R{r}
\def\Rmol{\R_{\textsc{\scriptsize m}}}
\def\Rgr{\R_{\textsc{\scriptsize g}}}
\def\Nmutr{N_\mu^{\rm tr}}

\runauthor{T.~Antoni et al. (KASCADE Collaboration) }
\runtitle{Air shower lateral distributions}

\begin{document}


\begin{frontmatter}

\title{Electron, Muon, and Hadron Lateral Distributions Measured
in Air-Showers by the KASCADE Experiment}

\author[KA-FZK]{T.~Antoni},
\author[KA-FZK]{W.\,D.~Apel},
\author[BU]{F.~Badea},
\author[KA-FZK]{K.~Bekk},
\author[KA-FZK]{K.~Bernl\"ohr\thanksref{nowKB}},\relax 
   \thanks[nowKB]{Now at: University of Hamburg, Hamburg.}
\author[KA-FZK,KA-Uni]{H.~Bl\"umer},
\author[KA-FZK]{E.~Bollmann},
\author[BU]{H.~Bozdog},
\author[BU]{I.\,M.~Brancus},
\author[YE]{A.~Chilingarian},
\author[KA-Uni]{K.~Daumiller},
\author[KA-FZK]{P.~Doll},
\author[KA-FZK]{J.~Engler},
\author[KA-FZK]{F.~Fe{\ss}ler},
\author[KA-FZK]{H.\,J.~Gils},
\author[KA-Uni]{R.~Glasstetter},
\author[KA-FZK]{R.~Haeusler},
\author[KA-FZK]{W.~Hafemann},
\author[KA-FZK]{A.~Haungs},
\author[KA-FZK]{D.~Heck},
\author[KA-FZK]{T.~Holst},
\author[KA-Uni]{J.\,R.~H\"orandel\thanksref{nowatChicago}},\relax 
   \thanks[nowatChicago]{Present address: 
      University of Chicago, Enrico Fermi Institute, Chicago, IL~60637.}%
\author[KA-Uni,KA-FZK]{K.-H.~Kampert},
\author[LZ-Dep]{J.~Kempa},
\author[KA-FZK]{H.\,O.~Klages},
\author[KA-Uni]{J.~Knapp\thanksref{nowatLeeds}},\relax 
   \thanks[nowatLeeds]{Now at: University of Leeds, Leeds LS2~9JT, U.K.}%
\author[KA-FZK,KA-Uni]{D.\,Martello},
\author[KA-FZK]{H.\,J.~Mathes},
\author[KA-FZK]{H.\,J.~Mayer},
\author[KA-FZK]{J.~Milke},
\author[KA-FZK]{D.~M\"uhlenberg},
\author[KA-FZK]{J.~Oehlschl\"ager},
\author[BU]{M.~Petcu},
\author[KA-FZK]{H.~Rebel},
\author[KA-FZK]{M.~Risse},
\author[KA-FZK]{M.~Roth},
\author[KA-FZK]{G.~Schatz},
\author[KA-Uni]{F.\,K.~Schmidt},
\author[KA-FZK]{T.~Thouw},
\author[KA-FZK]{H.~Ulrich},
\author[YE]{A.~Vardanyan},
\author[BU]{B.~Vulpescu},
\author[KA-Uni]{J.\,H.~Weber},
\author[KA-FZK]{J.~Wentz},
\author[KA-FZK]{T.~Wiegert},
\author[KA-FZK]{J.~Wochele},
\author[LZ-Sol]{J.~Zabierowski}%

\collab{(The KASCADE Collaboration)}

\address[KA-FZK]{Institut f\"ur Kernphysik, Forschungszentrum Karlsruhe,
      	     76021~Karlsruhe, Germany}
\address[BU]{National Institute of Physics and Nuclear Engineering, 
             7690~Bucharest, Romania}
\address[YE]{Cosmic Ray Division, Yerevan Physics Institute, 
             Yerevan~36, Armenia}
\address[KA-Uni]{Institut f\"ur Experimentelle Kernphysik, University of
             Karlsruhe, 76021~Karlsruhe, Germany}
\address[LZ-Dep]{Department of Experimental Physics, 
             University of Lodz, 90236~Lodz, Poland}
\address[LZ-Sol]{Soltan Institute for Nuclear Studies,
             90950~Lodz, Poland}
	     
\ifx AA
\makeatletter
\begingroup
  \global\newcount\c@sv@footnote
  \global\c@sv@footnote=\c@footnote     
  \output@glob@notes  
  \global\c@footnote=\c@sv@footnote     
  \global\t@glob@notes={}
\endgroup
\makeatother
\fi

\newpage

\begin{abstract}
\noindent Measurements of electron, muon, and hadron lateral
distributions of extensive air showers as recorded by the
KASCADE experiment are presented.  The data cover the energy
range from $5\times10^{14}$ eV up to almost $10^{17}$ eV and
extend from the inner core region to distances of 200~m.  The
electron and muon distributions are corrected for mutual
contaminations by taking into account the detector properties in
the experiment.  All distributions are well described by
NKG-functions.  The scale radii describing the electron and
hadron data best are $\simeq 30$~m and $\simeq 10$~m,
respectively.  We discuss the correlation between scale radii
and `age' parameter as well as their dependence on shower size,
zenith angle, and particle energy threshold.

\end{abstract}

\begin{keyword}
cosmic rays; air shower; lateral distribution
\PACS 96.40.Pq
\end{keyword}

\end{frontmatter}


\section { Introduction }
\label{sec:intro}

Since the detection of extensive air showers (EAS)
\cite{Auger-1939} lateral or radial density distributions
$\rho(r)$ of different kinds of particles produced in EAS have
been an ongoing target of experimental as well as theoretical
investigations.  There are a number of reasons why EAS lateral
distributions are of importance for the air shower phenomenon. 
The first and most important one is that from the number and
distribution of ground particles the energy and mass of the
primary particle can be deduced.  
While at least the energy reconstruction can be done rather crudely from
analytical considerations, more reliable algorithms need detailed air
shower simulations to relate the observables to primary energy and mass.
Perhaps trivial, although
experimentally very important, is the fact that in measurements
the shower particles are always sampled over a limited range of
core distances $r_1<r<r_2$ only -- in most cases with an area
coverage in this range not much exceeding one percent -- but
showers are often referred to in terms of integrated numbers of
particles:
\begin{equation}
  N = \int_{r_1}^{r_2} 2 \pi r \rho(r) dr. 
\label{eq:integral}
\end{equation}
The total particle numbers, $N$, for different kinds and energy
ranges of shower particles, are obtained by choosing $r_1=0$ and
$r_2=\infty$ and are traditionally used both as measures for the
primary energy in an individual experiment as well as a means for
comparison of different experiments.  Comparing experimental
results is generally hampered by different detector
characteristics like energy thresholds and discrimination of
particle types but also by different observation altitudes.  
These effects need to be corrected, e.g. using suitable air shower and
detector response simulations.

Detector effects are, in principle, minimized by disentangling the
lateral distributions (and thus also the total particle numbers) for
various kinds of particles.  For experiments using a
single type of detector this is generally not an option and resulting
lateral distributions correspond to some mixture of different particle
types, depending on the detection technique used as well as on
absorber thicknesses and thresholds applied.  Experiments with several
detector components optimized for different particle types still
measure a mixture of particles, but are able to disentangle various
types to a large extent.  The present paper follows this path to
obtain lateral distributions separately for the major charged particles
-- electrons, muons, and hadrons -- in EAS of primary energies from
$5\mal10^{14}$~eV up to almost $10^{17}$~eV.

Integrating from $r_1=0$ to $r_2=\infty$ implies an extrapolation
beyond the core distance ranges actually covered.  Deviations of
measured lateral distributions from the expected form, as a
consequence, result in systematic errors of the particle numbers
obtained.  Such systematic errors can be very much reduced by
using '{\em truncated\/}' particle numbers integrated only over
the experimentally covered range of core distances.  This
approach is introduced for muons in section
\ref{sec:shower-reco}.  The main drawback of such truncated
particle numbers is that comparisons of different experiments are
further complicated.

The lateral distribution functions also carry information on the
related particle physics and astrophysics.  Different hadronic
interaction models predict different lateral shapes. Hence, it
is possible to test available interaction models. 
Unfortunately, from a particle physicists point of view, the
expected lateral shape also depends on the mass of primary cosmic
rays.  Heavier primaries lead, on average, to a flatter
distribution.  Since the lateral distribution is only one in a
group of composition-sensitive observables measured with KASCADE,
no attempts are made to infer any mass parameter in the present
paper.  This will be addressed in a separate article.

Historically, choices of parametrizations of both electron and muon lateral
distributions were influenced very much by the seminal review of
Greisen \cite{Greisen-1960}.  Greisen described the lateral density
function ({\em LDF\/}) of the electromagnetic ({\em e.m.}) component
of EAS by
\begin{equation}
\rho_{\rm em}(\R,N_e)=\frac{0.4 N_e}{\Rmol^2} \;\;
  \Bigg(\frac{\Rmol}{\R}\Bigg)^{0.75}
  \Bigg(\frac{\Rmol}{r+\Rmol}\Bigg)^{3.25} 
  \Bigg(1+\frac{\R}{11.4\,\Rmol}\Bigg)
\label{eq:Greisen-em}
\end{equation}
over the core distance range from $r=5$~cm to $r=1500$~m and for
atmospheric depths of 537~g/cm$^2$ to 1800~g/cm$^2$.  The parameter
$N_e$ is the total number of electrons in the shower and $\Rmol$ is
the Moli\`ere radius.  The Moli\`ere unit, about 0.25 radiation
lengths in air, characterises the spread of low-energy electrons by
multiple scattering.

Greisen also noted that Eq.~\ref{eq:Greisen-em}, except
for the last factor, is a close approximation to the analytical
calculations for electromagnetic showers performed by Kamata and
Nishimura \cite{Kamata-1958} if a shower {\em age parameter\/} of
$s=1.25$ is assumed.  Greisen's approximation to the
Nishimura-Kamata functions for $0.5<s<1.5$ is referred to as the
NKG function:
\begin{equation}
\rho_{\rm NKG}(\R,s,N_e) = 
  \frac{N_e}{\Rmol^2} \;\;
  \frac{\Gamma(4.5-s)}{2\pi\Gamma(s)\Gamma(4.5-2s)} \;\;
  \Bigg(\frac{\R}{\Rmol}\Bigg)^{s-2} 
  \Bigg(1+\frac{\R}{\Rmol}\Bigg)^{s-4.5} .
\label{eq:NKG}
\end{equation}
This function, often used to describe the charged particle
lateral distribution, will in the following be applied
individually to electron, muon, and hadron 
distributions by choosing approriate sets of parameters
$(s,\Rmol)$.  For a comparison of the parametrization of Kamata
and Nishimura with the NKG function (Eq.~\ref{eq:NKG}) see
\cite{Nishimura-1967}.

Many experimental groups reported deviations of the e.m.\ LDF
from the NKG form which are most obvious at large core distances
\cite{Yoshida-1994,Glushkov-1997,Coy-1997}.  This may be related
to the problem that the NKG form has originally been formulated
for {\em zero} energy threshold of shower electrons in purely
electromagnetic showers and that higher moments of the NKG form
tend to diverge, depending of the age parameter $s$.  More
general forms were, for example, suggested by Hillas and Lapikens
\cite{Hillas-1977} and Capdevielle et al.\
\cite{Capdevielle-1977}.

Traditionally, the NKG form is used with a fixed value of $\Rmol$ and
a variable age parameter $s$.  Thus, the scale length is kept
constant while the shape of the LDF is assumed to be variable.  A
different LDF and a scaling relation were proposed by Lagutin et al.\
\cite{Lagutin-1997}, based on Monte Carlo calculations for pure
electromagnetic showers.  They proposed a normalized LDF $f(x)$ with
\begin{equation}
xf(x)=\exp(-3.63 - 1.89\ln x - 0.370\ln^2 x -0.0168\ln^3 x),
\label{eq:lagutin}
\end{equation}
independent of primary energy and age at least in the range
$0.05\le x\le25$, where $x$ is the core distance divided by a
scale radius, here the root mean square (rms) radius,
$\sqrt{\langle r^{2} \rangle}$, of the particle density at
ground.  Note that Eq.~\ref{eq:lagutin} has finite higher order
moments but is not useful for small $x$.

Greisen \cite{Greisen-1960} also suggested a functional form for the
muon LDF in EAS
\begin{equation}
\rho_{\mu}(\R,N_\mu) = \textrm{const.}\;\; N_\mu\;\; 
   \Bigg(\frac{\R}{\Rgr}\Bigg)^{-\beta}
   \Bigg(1+\frac{\R}{\Rgr}\Bigg)^{-2.5},
\label{eq:Greisen-mu}
\end{equation}
with $\beta = 0.75$, now referred to as the Greisen function.  In
the original form, which was based on a very limited number of
events, the Greisen radius $\Rgr$ is 320~m.  This form refers to
a minimum muon energy of 1~GeV but a more general form for the
1--20~GeV range was also quoted by Greisen.  Deviations of the
muon LDF from the Greisen form (Eq.~\ref{eq:Greisen-mu}) were
reported by several experimental groups, such as
\cite{Armitage-1987,Hayashida-1995,Aglietta-1997}.  Alternatives
were suggested by Linsley \cite{Linsley-1963} and by Hillas et
al.\ \cite{Hillas-1969}.  The KASCADE experiment allows to
scrutinize the muon LDF for different thresholds (0.23--2.4~GeV),
although only for core distances below 100--230~m.

For the hadronic (originally termed {\em nuclear interacting\/}
or just {\em N\/}) component, the LDF depends very much on the
hadron energy, with more energetic hadrons being more concentrated
near the shower core.  This was already pointed out by Greisen
\cite{Greisen-1960}.

Hadron lateral distributions were investigated by previous
experiments mainly close -- within 10~m or less -- to the shower
core.  Rather wide lateral distributions measured at Tien Shan
initiated speculations on rising mean values of transverse
momenta or on strongly rising cross-sections of jet production
\cite{Danilova-1985}.  At Chacaltaya hadrons far away from the
shower axis and with high values of $p_{T} \approx 2$~GeV/c have
been observed \cite{hasegawa65}.  Similar discrepancies between
observations and calculations were reported repeatedly, e.g.\ by
the Turku group, claiming an increase of large transverse
momentum processes with rising energy (see \cite{Arvela-1995} and
references therein).  Features suggesting strong changes in the
characteristics of hadronic interactions in the PeV range were
also claimed in \cite{Vatcha-1973a,Vatcha-1973b}.

Early parametrizations of the hadron LDF assumed a power law
form, as for instance shown in the review by Cocconi
\cite{Cocconi-1961}.  More recently, Maket-ANI reported agreement
with exponential forms within 5~m from the core
\cite{Ter-Antonian-1995}.  The Chacaltaya group found a NKG-like
function to fit their lateral distribution best for core
distances up to 10~m \cite{Ticona-1993a}.  The KASCADE hadron
calorimeter with its large detection area and large dynamic range
not only provides better statistics but also allows to extend the
hadron LDF analysis to much larger core distances than any 
previous experiment.

\section{The KASCADE experiment}

The KASCADE ({\em KArlsruhe Shower Core and Array DEtector\/})
experiment is located at Forschungszentrum Karlsruhe, Germany, at
an altitude of 110~m a.s.l.\ and has been described in detail in
\cite{Doll-1990b,AAA067.022.145}.  The experiment has three major
components: an array of electron and muon detectors, a central
detector mainly for hadron measurements but with substantial muon
detection areas, and a tunnel with streamer tube muon telescopes. 
Since the latter have only been completed at the time of writing
this article, no data from the muon tunnel are included in the
present analysis.

The KASCADE array covers an area of about 200$\times$200~m$^2$
and consists of 252 detector stations located on a square grid of
13~m separation.  These are organized in 16 clusters of 16
stations each, except for the inner four clusters where the
location of one station is blocked by the central detector.  The
stations contain two types of detectors, liquid scintillation
counters ($e/\gamma$ detectors) of 0.79~m$^2$ area each and 5~cm
thickness with little shielding above and plastic scintillators
of 0.81~m$^2$ area each and 3~cm thickness (muon detectors) below
a shielding of 10~cm lead and 4~cm steel.  The inner four
clusters are instrumented with four $e/\gamma$ detectors per
station but without muon detectors while the outer 12 clusters
house two $e/\gamma$ and four muon detectors per station.  For
both detector types the sum of photomultiplier signals, the
earliest time, and the hit pattern are recorded for all stations
fired.

The central hadron calorimeter is of the sampling type and has a
fiducial area of 16$\times$19~m$^2$.  A detailed description can
be found in \cite{Engler-1999}.  The energy is absorbed in an
iron stack and sampled in eight layers of liquid ionization
chambers with anode segments of 0.25$\times$0.25~m$^2$ (appr.\ 38\,500
channels).  The thickness of the iron slabs increases from 12 to
36~cm towards the deeper parts of the calorimeter, amounting to
154~cm in total.  The 8$^{\rm th}$ layer is located below an
additional concrete ceiling of 77~cm thickness.  On top, a 5~cm
lead layer filters off the soft electromagnetic component.  The
ionization chambers are read out by logarithmic amplifiers and 13
bit ADCs, achieving a dynamic range of $6\mal10^4$.  Signals
starting from single minimum ionizing muons up to energy deposits
of 10~GeV in a chamber are read out without saturation.  The
response curve of each channel is calibrated with a reference
capacitor coupled to the preamplifier, injecting known charges
into the electronics chain.

Below the 8$^{\rm th}$ calorimeter layer, two layers of multiwire
proportional chambers (MWPCs), vertically separated by 38 cm, are
used as muon detectors.  In total, 32 chambers are operated with
129~m$^2$ total area per layer.  Hits are registered on anode
wires and two layers of cathode strips at angles of $\pm34^\circ$
with respect to the wires.

A total of 456 plastic scintillation counters of $0.48 \times
0.95$~m$^2$ area each and 3~cm thickness are used within the
calorimeter (below 5~cm lead and 36~cm steel) to trigger the
calorimeter and the MWPCs.  They also serve as muon counters.  On
top of the calorimeter, an additional 50 such counters fill the
central gap of $e/\gamma$ detectors but their data are not
included in the present analysis.  A summary of the detector
components used in this article together with their most relevant
parameters is given in Table~\ref{tab:detectors}.

\begin{table}[t]
\caption{KASCADE detector components used in this analysis. 
Detection thresholds refer to particle energies above the 
absorber material of the detectors.}
\label{tab:detectors}
\def\secz{~$\times\sec\theta$}
\def\pscz{\phantom{\secz}}
\vspace*{3mm}
\begin{tabular}{lcccrc}
\hline
Detector & channels & separation & total area & threshold $E_{\rm kin}$~~ & for \\
\hline
array $e/\gamma$ & 252     & 13~m       & 490~m$^2$  &    5 MeV\pscz  & $e$ \\
array $\mu$      & 192     & 13~m       & 622~m$^2$  &  230 MeV\secz  & $\mu$ \\
trigger          & 456     &   ---      & 208~m$^2$  &  490 MeV\secz  & $\mu$ \\
MWPCs            & 26\,080 &   ---      & 129~m$^2$  &  2.4 GeV\secz  & $\mu$ \\
calorimeter      & 38\,368 &   ---      & 304~m$^2$  &   50 GeV\pscz  & hadrons \\
\hline
\end{tabular}
\end{table}

\section{Data analysis procedures}
\label{sec:analysis}

\subsection{Shower reconstruction}
\label{sec:shower-reco}

Shower parameters are reconstructed from the array data in a
procedure with three iterations.  In a first step core positions
are obtained by the centre of gravity ({\em COG\/}) of the
$e/\gamma$ detector signals.  The shower direction is determined
assuming a plane shower front.  In this first step, the electron
and muon shower sizes, $N_e$ and $N_\mu$, are obtained by summing
up the relevant detector signals of the $e/\gamma$ and muon
detectors, respectively, and multiplying them with a
core-position dependent geometrical weight factor.  Next, the
shower direction is obtained by fitting a conical shower front to
the recorded times of $e/\gamma$ detectors which are within 70~m
from the shower core.  The core position is fitted simultaneously
with the electron shower size and the electron lateral shape
parameter.  Next, the $e/\gamma$ detector signals are corrected for
expected contributions from particles other than electrons, and
muon detector signals are corrected for expected electromagnetic
and hadronic punch-through (see Section~\ref{sec:lecf}).  The
main differences between the second and third iteration are
improved corrections.  Signals largely inconsistent with those in
the neighbouring detectors or with expected fluctuations of
particle numbers, or signals more than some 200~ns off from the
shower front are discarded in this fit.  This greatly reduces the
impact of hadronic and electromagnetic punch-through on muon
signals.

In the shower reconstruction, both the electron and muon lateral
distributions are assumed to follow a NKG form with a Moli\`ere
radius of 89~m and 420~m, respectively.  For muons, the fit is
performed by considering only detector stations at core distances
between 40 and 200~m.  The {\em truncated muon size},
$N_{\mu}^{\rm tr}$, is then defined by integrating
Equation~\ref{eq:integral} in this range.  The lower integration
limit is imposed by severe punch-through near the shower axis and
the upper one is approximately the largest core distance of any
counter for showers with their cores inside KASCADE. The scale
radius of 420~m and the {\em muon age} parameter $s_\mu$ were
deduced from simulations.  The latter can not be fitted on a
shower-by-shower basis due to limited statistics.  Therefore, it
is derived from CORSIKA simulations \cite{CORSIKA} and
parametrized as a function of $N_e$.  The actual values are
obtained by fitting CORSIKA muon density distributions
individually for proton and iron induced showers and taking the
mean value of both parametrizations.  This leads to a mass
dependent systematic error in the reconstruction of
$N_{\mu}^{tr}$ in a range of up to 5 $\%$ but yields a more
robust result than trying to fit also $s_{\mu}$ on a
shower-by-shower basis.

The actual muon LDF is known to deviate from the assumed NKG form
outside the fit range.  Within the limited range accessible to
the experiment, the NKG form is nevertheless, on a
shower-by-shower basis, as good as the Greisen function or any
other form with suitably adapted parameters.  Simulations show
that $\Nmutr$ provides a very good estimate of the primary
energy, almost independent of the primary particle mass
\cite{Weber-1997}.

The adopted electron LDF is also known to deviate from
experimental data at large core distances, even though the age
parameter is fitted on a shower-by-shower basis for core
distances of 10--200~m.  This results in an underestimate of
$N_e$ up to 5--8\% for simulated showers.

The core position can be reconstructed with an uncertainty of
about 3~m at 1~PeV, and the accuracy is typically better than 1~m
for showers above 4~PeV if the core is located well inside the
array.  In order to use data of best quality, the analysis of
average lateral distributions in this paper is restricted to
showers with core positions within 91~m from the centre of the
array.  The angular resolution for such showers above 1~PeV is
about 0.4$^\circ$ (68\,\% C.L.).  Statistical sampling errors on
$N_e$ ($N_\mu^{\rm tr}$) improve from about 10\% (20\%) at 1~PeV
to about 3\% (10\%) at 10~PeV.


\subsection{Particle numbers in array detectors}
\label{sec:ana-array}
\label{sec:lecf}

The signal analysis of the $e/\gamma$ detectors takes account
of muons by subtracting their expected energy deposits from the
measured energy.  In the same way, the expected $e/\gamma$ and
hadron punch-through contributions are subtracted from the energy
deposit in the muon detectors.  Since both are mutually related,
an iterative procedure is applied.  Finally, the number of
particles hitting a detector is estimated by dividing the
remaining energy deposit by the expected energy deposit per
particle.  There are no means to discriminate against hadrons in
the array detectors.

\begin{figure}
\hc{\epsfig{file=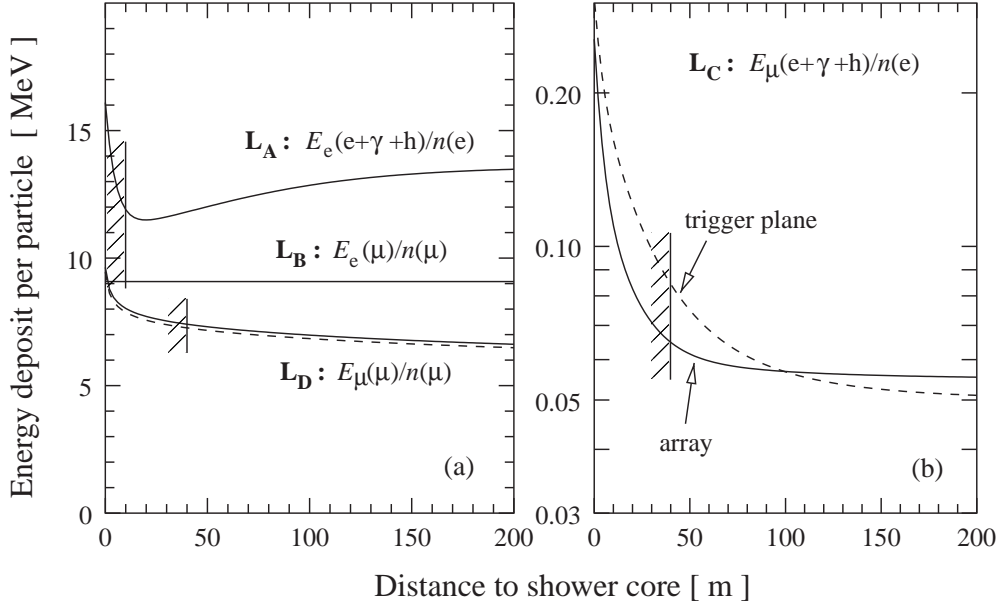,width=0.95\hsize}}
\caption[Lateral energy correction functions]{Parametrized
lateral energy correction functions as used in the analysis of
average LDFs.  Left: $L_{\rm A}$ and $L_{\rm B}$ for deposits in
the array $e$/$\gamma$ counters (see text) and $L_{\rm D}$ for
muons in array muon counters (solid line) and trigger plane
counters (dashed line). Right: Punch-through corrections
(evaluated for $N_e=10^5$) for array muon and trigger plane
counters, corresponding to final likelihood cuts used to reject
counters with far too large energy deposits.
Correction functions left of the hatched lines
are only shown for illustration but not used in the shower 
reconstruction.}
\label{fig:lecf}
\end{figure}

In total, the procedure requires four {\em lateral energy
correction functions} (LECFs) which are, in the most general
case, a function of core distance, zenith angle, and shower size. 
Actually, there is a weak dependence on the cosmic-ray mass
composition, too.  Hence, mean values of the LECFs for simulated
proton and iron primaries are used.  All LECFs were obtained from
CORSIKA simulations of EAS (using the QGSJET model \cite{qgsjet})
followed by detailed detector simulations (based on the GEANT
package for detector simulation \cite{GEANT}).  The same event
reconstruction is applied to simulated as to experimental data. 
As a result, the energy deposits can be related to the numbers
and types of particles hitting the detectors.  For the analysis
of average lateral distributions special refined LECF functional
forms are used (see Fig.\ \ref{fig:lecf}).  They do not change
fundamental shower parameters like $N_e$ or $N_\mu^{\rm tr}$, but
improve the reproduction of the average lateral distribution of
simulated showers.  LDFs obtained after detector simulation and
shower reconstruction closely match LDFs of the relevant particle
type and energy threshold before detector simulation.

The following correction functions are used: $L_{\rm A}$ is the
average sum of energy deposits by electrons, gammas, and hadrons
in the $e/\gamma$ counters normalized to the number of electrons 
above 5~MeV kinetic energy, $n(e)$, hitting the counter;
$L_{\rm B}$ is the average energy deposit by muons in the
$e/\gamma$ counters per muon; $L_{\rm C}$ is the average
punch-through energy deposit of electrons, gammas, and hadrons in
the muon counters per electron; $L_{\rm D}$ is, in turn, the
average deposit of muons in the muon counters per muon.  The
average energy deposits $E_{e\gamma}$ and $E_\mu$ in $e/\gamma$
and muon detectors, respectively, are:
$$ E_{e\gamma} = A_{e\gamma} (L_{\rm A} \rho_e + L_{\rm B} \rho_\mu) $$
and
$$ E_\mu = A_\mu (L_{\rm C} \rho_e + L_{\rm D} \rho_\mu), $$
with $A_{e\gamma}$ and $A_\mu$ being the areas of  the
detectors and $\rho_e$ and $\rho_\mu$ the particle densities. 
Parametrized correction functions are given in Fig.~1.
The rise of $L_{\rm A}$ at small and
large core distances is due to hadrons and gammas, respectively. 
$L_{\rm D}$ rises at small distances because of energetic knock-on
electrons released by high-energy muons in the absorber material
above the muon counters.

$L_{\rm A}$ and $L_{\rm B}$ are adjusted for the core distance range
10--200~m, and $L_{\rm C}$ and $L_{\rm D}$ for the range 40--200~m.
For core distances below 20--40~m the electromagnetic and
hadronic punch-through, even on average, exceeds the energy
deposit by muons in the muon counters.  Despite the reduction of
'spikes' and correction for other punch-through, systematic
errors on the muon LDF due to punch-through remain significant
below 40~m core distance.  For this reason, the array muon
analysis was restricted to core distances farther than 40~m.

The expected numbers of the `wrong' particle types in the second
and third iteration are derived from the fitted lateral
distributions of the previous step.  Therefore, to some extent
the derived shower sizes depend on the assumed shapes.  The
electron size $N_e$, in particular, depends on the extrapolation
of the assumed electron LDF to radii below 10~m and above 200~m,
as well as on the assumed muon LDF at radii below 40~m -- the
latter resulting in about 2\% systematic error.

After shower reconstruction lateral distributions of average
energy deposits are obtained for various bins of shower size and
zenith angle.  A similar procedure to the one described above is
applied to correct for the contribution of other particle types,
but now on the {\em average} and not individual energy deposits. 
Therefore, refined LECFs are used to get average particle density
distributions.  Again, the corrections for muons in the
$e/\gamma$ detectors and for electron punch-through into the muon
detectors are calculated iteratively, by using LDFs from the
preceding iteration.  In each iteration, both electron and muon
LDFs are fitted and the slope $s_\mu$ of the muon LDF is no
longer parametrized (see Sec.~\ref{sec:shower-reco}) but is
fitted as well.  In these fits, we take into account not only the
statistics of hits but also the uncertainty expected in the
punch-through correction and we consider effects caused by the
experimental resolution in determining the core position.


\subsection{Muon numbers in the trigger plane}

The trigger plane has a muon energy threshold of 490~MeV and,
therefore, less electromagnetic punch-through than the
array muon detectors.  On the other hand, the dynamic range of
signals is smaller and hadronic punch-through is more significant
because of the outset of cascading in the absorber material
above.  The effect of hadronic punch-through is reduced by
rejecting detector elements within a distance of 1~m to
identified high-energy hadrons (typically above 50 GeV).  Further
reduction of hadronic and electromagnetic punch-through is
achieved by rejecting counters with energy deposits inconsistent
with the expected numbers of muons, accounting for statistical
number fluctuations and for fluctuations in the energy loss of
muons.

Conversion of average energy deposits into particle numbers in
the trigger plane closely follows the procedure outlined for the
array detectors.  LECFs for trigger plane counters were derived
from simulations -- with the same selection criteria applied as
for experimental data.  Most notably, the punch-through
correction is different for array detectors
(Fig.~\ref{fig:lecf}b), but the different composition of
materials above the muon counters also results in slightly
different effective energy loss distributions
(Fig.~\ref{fig:lecf}a).


\subsection{Reconstruction in the MWPCs}

Hits of single muons in the MWPCs are characterised by signals on
one or a few anode wires and an average of 3.5 neighbouring
cathode strips on each side.  Hit reconstruction requires that
the intersection of the two cathode signals coincides with an
anode signal.  At low particle densities this reconstruction
achieves a good efficiency and spatial resolutions are 1.4~cm
along wires and 0.7~cm perpendicular.  At high densities of about
5 muons/m$^2$, signals of several hits start to overlap and
ambiguities arise in the reconstruction.

Muon tracks are reconstructed from pairs of hits in the two
detector layers.  Accepted tracks are required to be in
reasonable agreement with the shower direction ($\Delta\theta \le
15^{\circ}$; $\Delta\phi \le 45^{\circ}$ if $\theta \ge
10^{\circ}$), which effectively resolves ambiguous hits. 
Systematic uncertainties in calculating efficiencies are reduced
by discarding those muons which, according to the shower
direction, could be observed in one layer of the MWPCs only. 
Thereby, hits near the edges of the chambers are rejected. 
Furthermore, muons entering from the sides of the building in
inclined showers would have a lower than nominal energy
threshold.  To compensate for this effect, only those areas of
the MWPCs are used where muons parallel to the shower directions
have penetrated the entire iron absorber of the calorimeter. 
Geometric and reconstruction efficiencies were obtained from the
shower and detector simulation chain followed by the normal
reconstruction procedure.  The particle detection efficiency in
the MWPCs itself is derived continuously by using muons observed
in the trigger layer and the other of the two MWPC layers and is
typically 98\,\%.  All efficiencies are accounted for in the
lateral distributions.


\subsection{Hadron reconstruction}
\label{sec:hadrec}

Briefly, the algorithm for pattern recognition of hadrons in the
eight layers of ionization chambers proceeds as follows
\cite{Engler-1999}: Clusters of energy are searched to line up in
the calorimeter and to form a track in different layers from
which an approximate angle of incidence can be inferred.  Then,
patterns of cascades are searched for in the deeper layers. 
Going upwards in the calorimeter, clusters are formed from the
remaining energy and are lined up to showers according to the
direction already found.  The reconstruction efficiency for
isolated hadrons is 70\% at 50~GeV and reaches nearly 100\% at
100~GeV.

Hadron energies are reconstructed from the sum of calibrated
signals in layers 2--8, weighting each layer by the relative
amount of preceding absorber.  The uppermost layer is not used
for the energy determination to avoid distortions by
electromagnetic punch-through.  The weighted signal sum is
converted to energies by a function derived from detector
simulations based on the GEANT package.  The energy resolution is
rather constant, slowly improving from 20\% at 100~GeV to 10\% at
10~TeV.

Due to the fine lateral segmentation of 25~cm, the minimal
distance to separate two equal-energy hadrons with a 50\%
probability amounts to 40~cm.  This causes the reconstructed
hadron number density to flatten off at about 1.5 hadrons/m$^2$. 
The reconstructed hadron energy density, on the other hand, is
not affected by this saturation \cite{Engler-1999}.  Radiation
from high-energy muons can mimic hadrons.  However, their
reconstructed energies are much lower than those of the actual
muon, typically by a factor of 10.  Simulations show that a
1~TeV muon is identified as a hadron with a probability of
about~1\% \cite{mielke-96}.

\section{Electron lateral distribution}
\label{sec:electron}

\begin{figure}
\hc{\epsfig{file=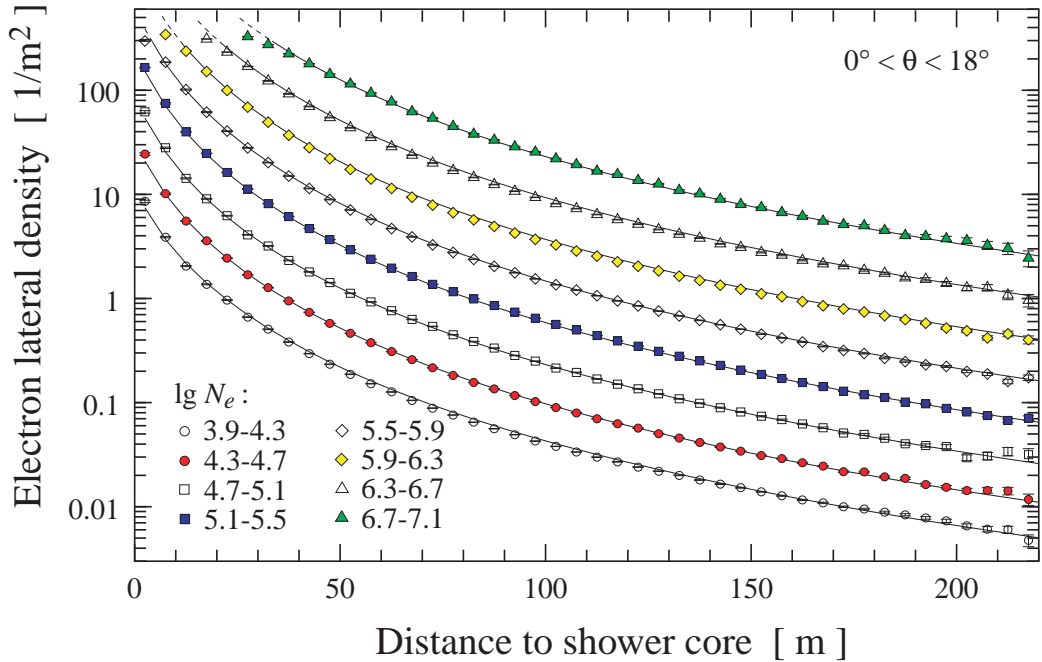,width=\hsize}}
\caption[Lateral distributions]{Lateral distributions of electrons
above a 5~MeV kinetic energy for zenith angles
below 18$^\circ$.  The lines show NKG functions of fixed age
parameter $s=1.65$ but varying scale radius $r_{e}$ (see text).}
\label{fig:e-lat-c2}
\end{figure}

Average LDFs of electrons have been reconstructed for shower
sizes from less than $10^4$ to more than $10^7$ electrons. 
Contributions of hadrons, muons, and gammas to energy deposits in
the $e$/$\gamma$ detectors were corrected for by the procedure
outlined in Section~\ref{sec:ana-array}.  Resulting lateral
distributions for electrons above 5~MeV kinetic energy are
presented in Fig.~\ref{fig:e-lat-c2}.  NKG functions fit the data
quite well and are represented by lines.  Reduced $\chi^{2}$
values are $\simeq$ 1--3 with only the lowest $N_{e}$ bin being
worse by about a factor of 2.  Note that saturation effects in
{\em average} lateral distributions become relevant at electron
densities of approximately 300~m$^{-2}$ although individual
counters have a dynamic range of up to 600~m$^{-2}$.  The effect
is attributed to shower-by-shower fluctuations and it does not
influence the measurement in individual events.

\begin{figure}
\hc{\epsfig{file=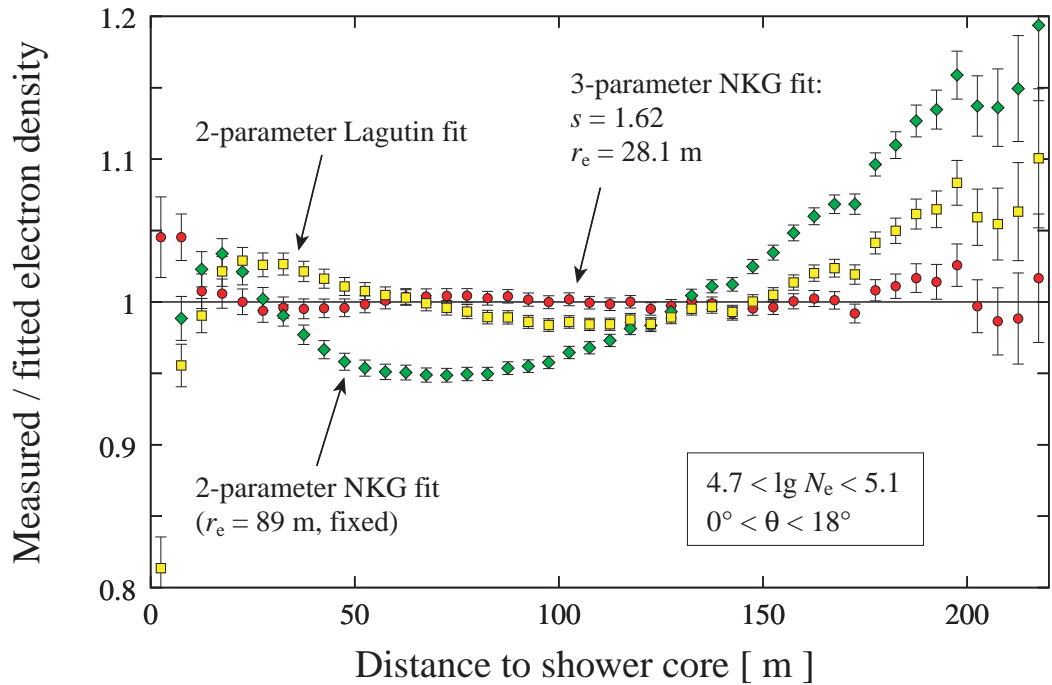,width=\hsize}}
\caption[Fits to the lateral distribution]{Residuals in the ratio of
measured over fitted average LDF, with the NKG function (3-parameter
fit, circles: $N_e$, $r_{e}$ and $s$ free; 2-parameter fit, diamonds:
$r_{e}$ fixed at 89~m) and the function (\ref{eq:lagutin}) proposed by
Lagutin \cite{Lagutin-1997} (squares; $N_e$ and rms-radius
fitted).}
\label{fig:fit-quality}
\end{figure}

\begin{figure}
\vspace*{5cm}
\caption[Correlation of $r_{e}$ and $s$]{Logarithm of the reduced
$\chi^2$ when fitting the KASCADE average electron lateral distribution
by the NKG function, 
with $r_{e}$ and $s$ varied over a wide range (only $N_e$ fitted).}
\label{fig:cntr-e}
\end{figure}

Deviations of the experimental LDF from the NKG function
(Eq.~\ref{eq:NKG}) have been discussed in the literature
frequently (see Section \ref{sec:intro}) and are subject to more
detailed studies presented below.  It turns out that the NKG
function {\em can\/} describe the KASCADE electron LDF over the
core distance range 10--200~m surprisingly well -- but the best
agreement is achieved with parameters far away from the
conventional assumption of $\Rmol\approx80$~m.  When fitting
$N_e$, $\Rmol$, and $s$ simultaneously, the measured LDFs can be
reproduced at the 1\% level for $\Rmol \approx20$--30~m and
$s\approx1.6$--1.8.  The actual values depend on shower size and
zenith angle.  In order to avoid confusions with the original
Moli\`ere radius, $\Rmol$, we will call this fit parameter of the
electron lateral distributions $r_{e}$ in the following. 
Two-parameter fits with conventional $r_{e}$ are substantially
worse, as can be seen in Fig.~\ref{fig:fit-quality}.  Similar
results have been found in $\gamma$-shower calculations by Hillas
and Lapikens \cite{Hillas-1977}.  In the fits, we find a strong
correlation between $r_{e}$ and $s$.  A corresponding parameter
map has been generated by scanning $(s,r_{e})$ and fitting
$N_{e}$.  The reduced $\chi^2$-values are about one
and are plotted as black areas in Fig.\,\ref{fig:cntr-e}.  It
seems worth noting that the optimal value of $r_{e}$ tends to
decrease and of $s$ to increase, both with increasing shower size
and with increasing zenith angle.  This variation does not
exactly follow the ridge shown in the $r_{e}$--$s$ plane of
Fig.\,\ref{fig:cntr-e}.

\begin{figure}
\hc{\epsfig{file=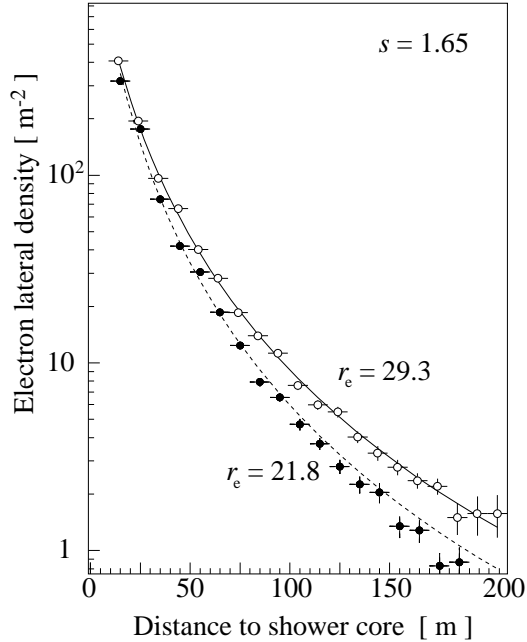,width=7cm}}
\caption[single events]{Lateral distributions of two EAS events
as measured by KASCADE.  Both distributions are approximated by
NKG functions of fixed age parameter $s = 1.65$. The fit parameters
are $r_{e} = 29.3$~m, $N_{e} = 2.3\cdot10^{6}$ (full line),
and $r_{e} = 21.8$~m $N_{e} = 2.0\cdot10^{6}$ (dashed line).
The reconstructed zenith angles are $\theta=26^{\circ}$
and $\theta=30^{\circ}$, respectively.}
\label{fig:elat-single}
\end{figure}

At the level of accuracy possible in individual showers, the
correlation between $r_{e}$ and $s$ entail highly ambiguous
values if $r_{e}$ and $s$ are fitted simultaneously.  Therefore,
$r_{e}$ is usually fixed and the steepness of the lateral
distributions is quantified by the fit parameter $s$.  A problem
mostly relevant to NKG functions with small scale radii $r_{e}
\simeq 25$~m is, that upwards fluctuations of the large $s$
parameter easily lead to ill defined shower sizes.  CORSIKA
simulatios show, that this problem can be circumvented if
$s$ is fixed and $r_{e}$ fitted, instead.  An example of
experimental electron lateral distributions of events with
similar shower size $N_{e}$ and zenith angle is presented in
Fig.~\ref{fig:elat-single}.  The difference in shape is clearly
visible and well accounted for by the different $r_{e}$
parameters of the fitted NKG functions.  The sensitivity of
$r_{e}$ to the primary mass will be subject of a forthcoming
publication.

Since evaluation of $N_e$ involves an extrapolation of the LDF
beyond the fiducial range, fitting with non optimal $r_{e}$
causes systematic errors of the shower size obtained.  In case of
KASCADE, variations of the Moli\`ere radius in the range
$25\textrm{\ m}<r_{e}<89\textrm{\ m}$ change $N_e$ by up to 5\%. 
Using a {\em truncated electron size\/} in analogy to $N_\mu^{\rm
tr}$, no such extrapolation would be required.  Since the
required correction is fairly small -- in contrast to the $N_\mu$
case -- we keep using a total electron shower size.

\begin{figure}
\hc{\epsfig{file=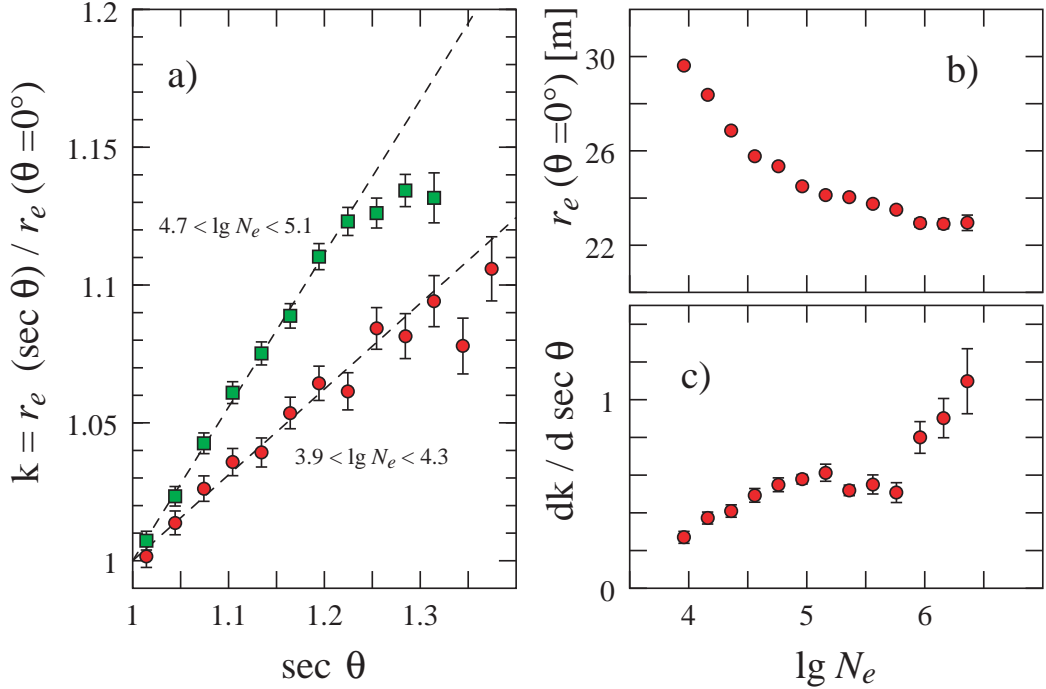,width=\hsize}}
\caption[]{a)~Shower scale radius $r_{e}$ as a function $\sec\theta$,
normalised to the scale radius for vertical showers.
The lateral distributions were fitted by NKG functions
with $s=1.65$ fixed.  Only showers with
$3.9 <\lg N_e <4.3$ and $4.7 <\lg N_e <5.1$ have been considered.
The dashed lines represent linear fits in the range $1< \sec \theta <1.25$.
b)~Radius parameter $r_{e}$ for vertical incidence.
c)~Slope of the linear function fitted as shown in a) as a
   function of shower size.}
\label{fig:e-zenith}
\end{figure}

Figure~\ref{fig:fit-quality} shows, in addition to the NKG
function, also residuals for a fit with the LDF of
Eq.~\ref{eq:lagutin} proposed by Lagutin et al.  The rather
moderate description of the experimental data is not particularly
surprising, since Eq.~\ref{eq:lagutin} was developed for purely
electromagnetic showers.  However, this form has a much more
reasonable behaviour at large core distances than the NKG form. 
The rms-radius
%
%
calculated directly from the experimental data and used by
Lagutin et al.\ as the scale radius, would be diverging when
evaluated analytically from our best-fit NKG form.  Lagutin et
al.\ also claimed that the shape of the LDF remains unchanged
when the radius is expressed in units of a variable scale radius. 
We tested this scaling hypothesis by fitting electron lateral
distributions of different shower sizes and zenith angles to the
NKG form and fixing the age parameter to $s=1.65$, a value which
provided the best overall fit for all data considered.  Again,
the reduced $\chi^{2}$ is in the range 1--4 with slightly better
fits obtained for larger shower sizes.  No siginificant
dependence on zenith angle is observed.  The parameter $r_{e}$ of
the fits can then be considered the variable scale radius.  As
shown in Fig.~\ref{fig:e-lat-c2}, all data can be reproduced
rather well -- with the largest deviations seen for shower sizes
$3.9 < \lg N_e < 4.3$.  This probably is due to selection effects
at threshold.  For small zenith angles, $r_{e}$ varies from about
30~m at $N_e=10^4$ to 24~m at $N_e=10^6$, with little change
beyond that size.  A comparison with results from 3-parameter NKG
fits to the electron LDF reveals that the scaling assumption does
indeed reproduce all our electron data rather well, but residuals
of up to $\simeq 5$\,\% also demonstrate significant deviations
from perfect scaling.

Apart from the scaling for different shower sizes it is
particularly illustrative to see a change of the scale radius as
a function of zenith angle.  To be independent of the scale
chosen and the precise form of the LDF used, the scale radius
shown in Fig.~\ref{fig:e-zenith}a is normalized to that for
vertical showers.  Using the Lagutin function instead of NKG,
consistent results are obtained.  A linear relation between scale
radius and the secant of the zenith angle $\theta$ is obvious and
is mainly a result of the increasing distance between the
detector and the shower maximum.  Electrons are being scattered
away from the shower axis as the shower has to penetrate a larger
air mass.  The slope in this relation can, in fact, be used to
infer the average depth of shower maximum -- although additional
corrections based on simulations have to be applied.  The
increase of the depth of shower maximum with increasing shower
size, and thus energy, results in a decrease of the scale radius
as seen in Fig.~\ref{fig:e-zenith}b.  Also, the slope of the
normalised scale radius versus $\sec\theta$ rises with shower
size (see Fig.~\ref{fig:e-zenith}c).  An exception from the
otherwise monotonic change is apparent at shower sizes
corresponding to the knee in the flux spectrum, which we observe
at $\lg N_{e} = 5.7$ \cite{Glasstetter99}.
A quantitative analysis of the phenomenon in terms of possible
change of the chemical composition is beyond the scope of this
paper.

\section{Muon lateral distributions}
\label{sec:muon}

The KASCADE experiment measures lateral distributions of muons
for three different energy thresholds
(Table~\ref{tab:detectors}).  In the following, we group the
showers in bins of truncated muon numbers $\Nmutr$. 
Punch-through and efficiency corrections are applied as described
in Section~\ref{sec:analysis}.  Ranges of core distances for the
different muon energy thresholds are limited by uncertainties in
the punch-through corrections at small core distances and by the
geometry of the KASCADE detector array.  Since $N_e/\Nmutr$ rises
with shower size, the impact of punch-through corrections becomes
more severe at higher energies and the minimum core distances
have to be increased correspondingly.  For showers with cores
inside KASCADE, the upper limit is about 220~m for array
detectors and 100~m for central detector components.

\begin{figure}
\hc{\epsfig{file=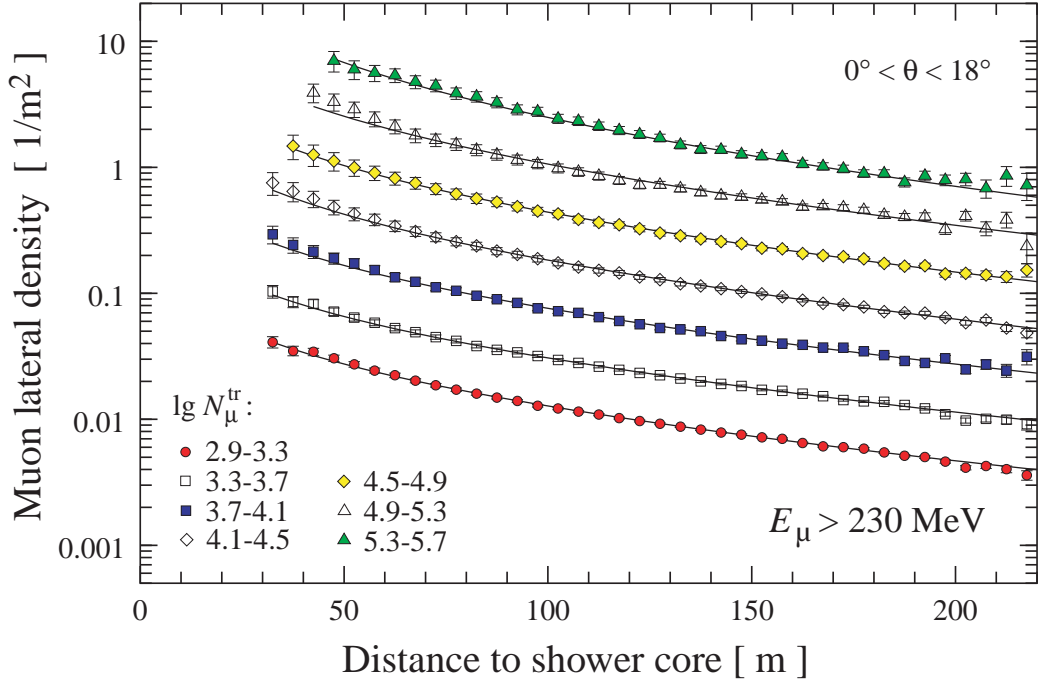,width=\hsize}}
\caption[]{Lateral distribution of muons above 230~MeV kinetic energy,
measured with the array detectors.  The lines indicate NKG
functions fitted to the data.  Error bars are of statistical 
nature including an uncertainty of 10\% on the punch-through
correction applied.}
\label{fig:mu-lat-array}
\end{figure}

\begin{figure}
\hc{\epsfig{file=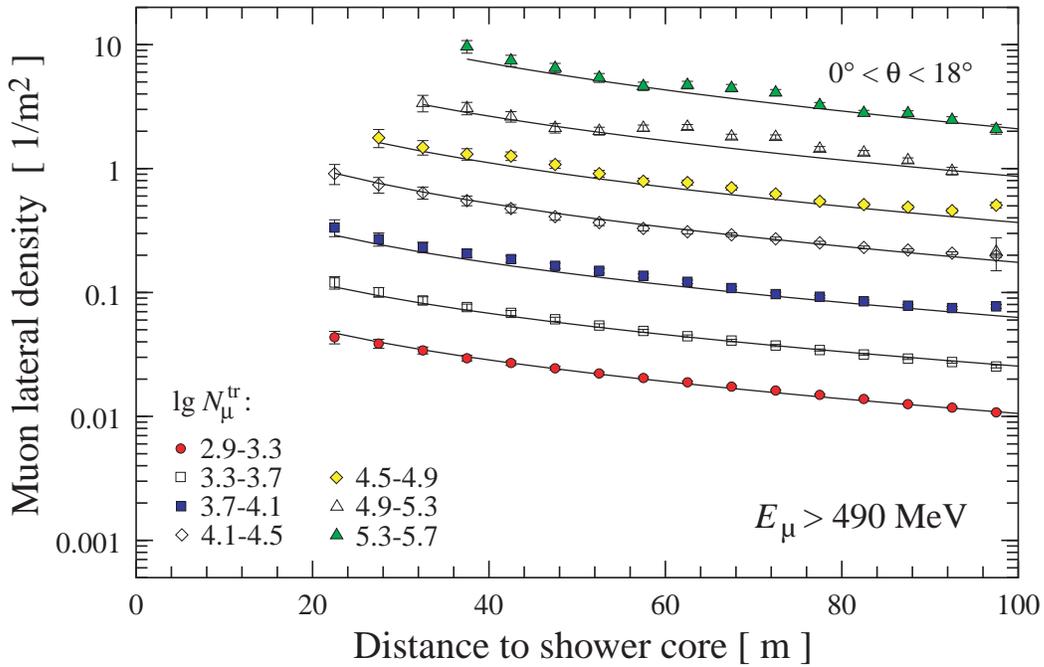,width=\hsize}}
\caption[]{Lateral distribution of muons above 490~MeV kinetic energy,
as measured with the trigger plane. Data are binned according to 
$N_{\mu}^{\rm tr}$ as measured by the array and the lines
represent NKG fits to the data.}
\label{fig:mu-lat-trigger}
\end{figure}

As is well known, the muon LDF at sea level and in the energy
range considered in this work is much flatter and typically an
order of magnitude lower than the electron LDF in our range of
core distances.  Figure~\ref{fig:mu-lat-array} presents average
muon lateral distributions with an energy threshold of 230 MeV.
NKG functions with $r_{\mu}=420$~m are superimposed as dashed
lines and typically fit the data to better than 5\%.  The Greisen
function (Eq.~\ref{eq:Greisen-mu}) or the forms suggested by
Linsley \cite{Linsley-1963} and by Hillas et al.\
\cite{Hillas-1969} do not -- within the small range of accessible
core distances -- provide a substantially better description of
the data.  The rather unconventional application of the NKG form
in fitting the muon lateral distribution for individual showers
(see Section \ref{sec:analysis}) is, therefore, not expected to
affect the quality of the $\Nmutr$ measurement.  As in the case
of electrons (see Fig.~\ref{fig:cntr-e}), the $s$ and $r_{\mu}$
values in NKG fits and also the $\beta$ and $\Rgr$ values in
Greisen fits are highly correlated.  While the total number of
muons $N_\mu$ is affected by this ambiguity of the scale radius,
$\Nmutr$ is not, because no extrapolation beyond the fiducial
core distance range is performed.

\begin{figure}
\hc{\epsfig{file=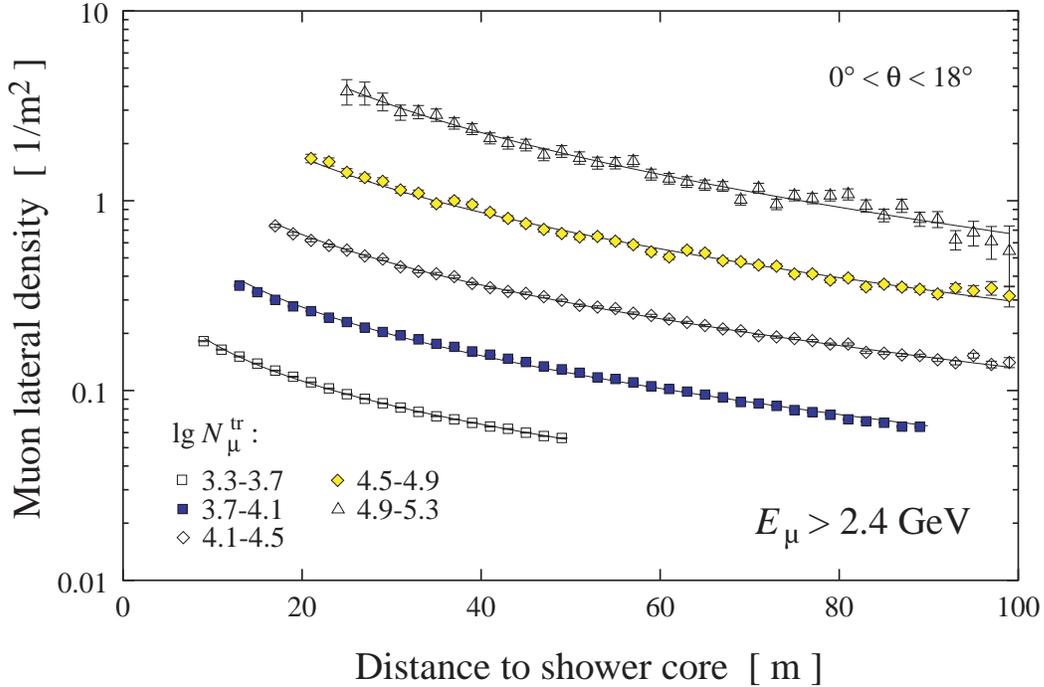,width=\hsize}}
\caption[]{Lateral distribution of muons above 2.4~GeV kinetic energy.
Error bars are statistical only. Data are binned according to 
$N_{\mu}^{\rm tr}$ as measured by the array and the lines
represent NKG fits to the data.}
\label{fig:mu-lat-mwpc}
\end{figure}

Muon density distributions above a threshold of 490 MeV as obtained
with the trigger plane detectors are presented in
Fig.~\ref{fig:mu-lat-trigger}.  Again, they are equally well fitted
by NKG as by Greisen functions.  Due to the smaller detector area,
statistical errors are larger than for the array muon LDF.
Nevertheless, the same range of $\Nmutr$ is covered, allowing to
compare both muon LDFs.  Apart from threshold effects this comparison
can serve as an additional check for any systematics, for example due
to punch-through corrections or cuts applied which are quite different
in both cases.

\begin{figure}
\hc{\epsfig{file=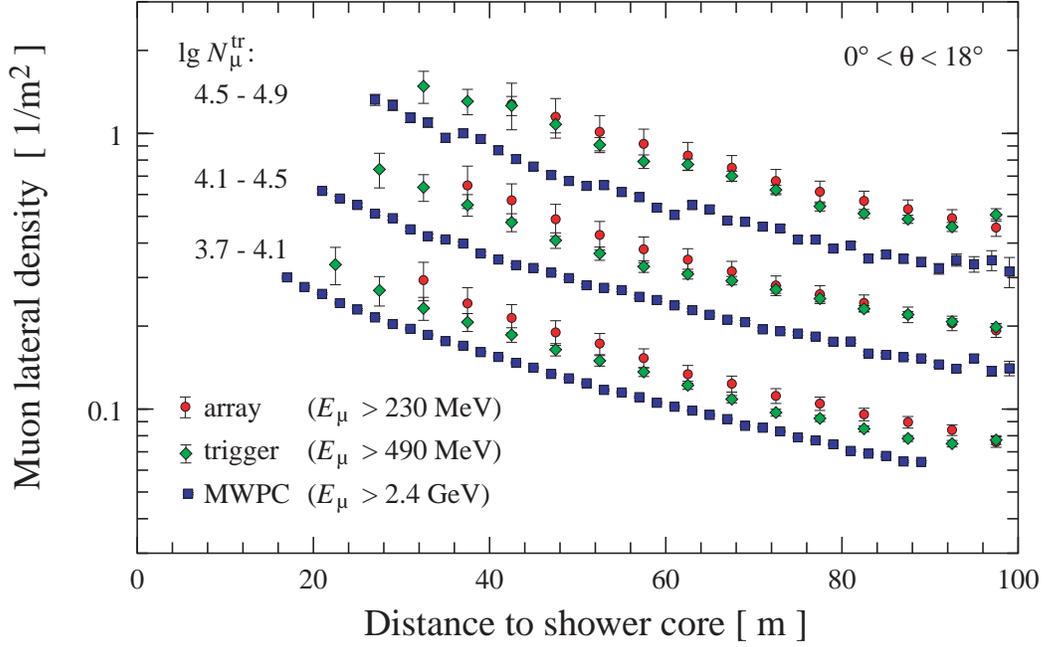,width=\hsize}}
\caption[]{Comparison of lateral distributions as measured with
the different KASCADE detector components, for three different
intervals of $\Nmutr$.}
\label{fig:mu-compare}
\end{figure}

Muons with energies above 2.4 GeV are measured with the MWPC
system and their LDF is presented in Fig.\,\ref{fig:mu-lat-mwpc}. 
Since these detectors are triggered only by the scintillators of
the trigger plane and not by the array stations, full efficiency
is reached only above the trigger threshold presently set at 7 
counters in the trigger plane, i.e.\ at $\rho_{\mu} \simeq
0.04$~m$^{-2}$.  The chambers identify a muon as a track and no
punch-through correction is applied.  The lower core distance
limit is applied mainly because of hadronic punch-through in EAS
cores.

\begin{figure}
\hc{\epsfig{file=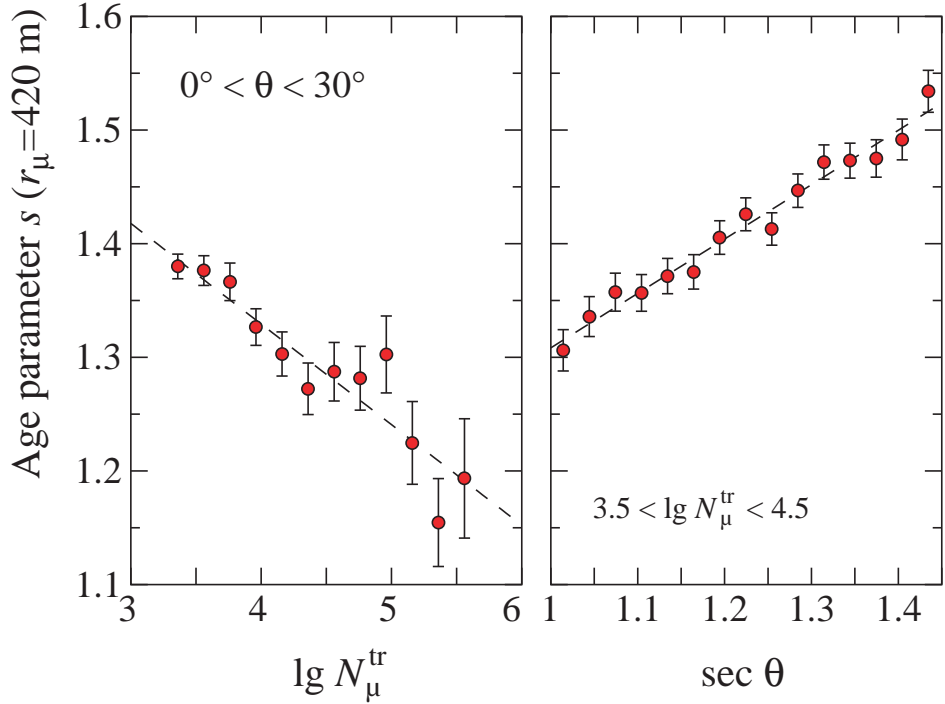,width=0.9\hsize}}
\caption[]{The age parameter $s$ in NKG fits with
fixed $r_{\mu}=420$~m for muons of 230 MeV threshold. a)~Dependence on
$\Nmutr$, for zenith angles below 30$^\circ$.  b)~Dependence on
the secant of zenith angle $\theta$, for $3.5< \lg \Nmutr <
4.5$. The dashed straight lines are drawn to guide the eye.}
\label{fig:mu-age-size-secz}
\end{figure}

The lateral distributions obtained for the different thresholds
are compared in Fig.~\ref{fig:mu-compare}.  As expected, the muon
density decreases with increasing threshold.  The drop of about
10\% between 230 and 490~MeV and of about 50\% between 230~MeV
and 2.4~GeV is nearly independent of primary energy and only
weakly dependent on core distance.  

For all thresholds, the muon LDF flattens with increasing zenith
angle and steepens with increasing shower size (see
\cite{Haungs-1999}).  This is illustrated in
Fig.~\ref{fig:mu-age-size-secz} for muons above 230~MeV, where
the age parameter $s$ in NKG functions fitted with
a fixed $r_{\mu}=420$~m is shown.  The observed effect is
comparable to that of the electron LDF. With increasing primary
energy, i.e.\ rising $\Nmutr$, the shower penetrates deeper into the
atmosphere resulting in steeper lateral distributions.  With
increasing zenith angle, in contrast, the shower maximum
recedes from the experiment resulting in correspondingly
flatter distributions.  This is partly compensated by a
harder muon spectrum which is due to longer decay path lengths.

\section{Hadron lateral distribution}
\label{sec:hadron}

\begin{figure}
\epsfig{file=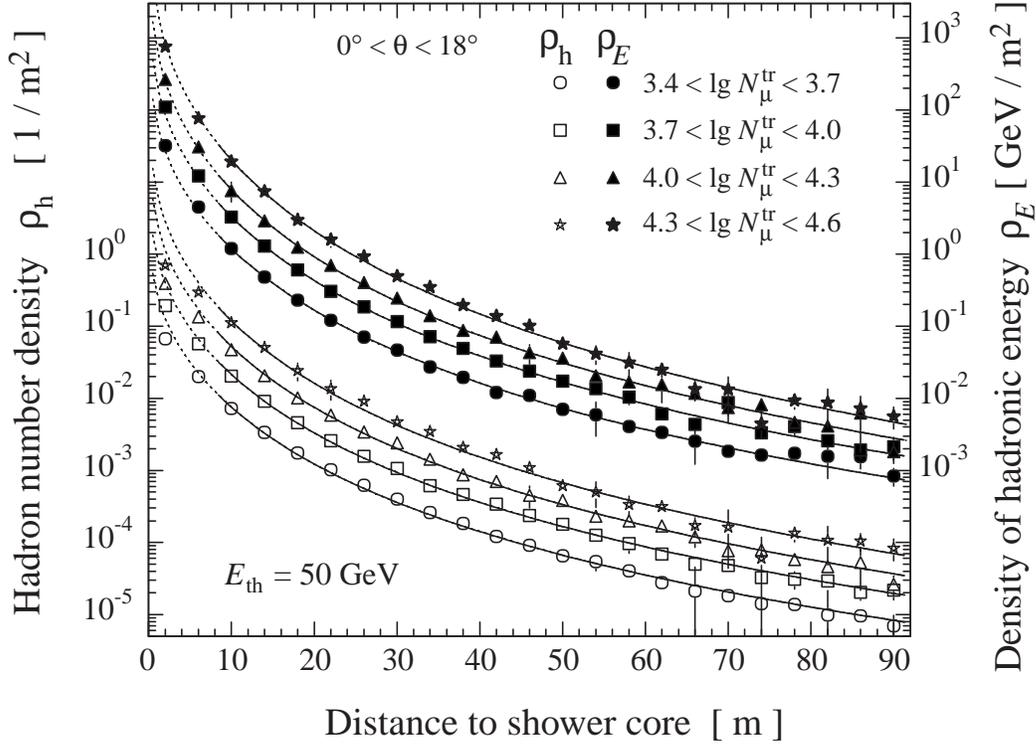,width=\textwidth}
\caption{Density of hadron number (left scale, open symbols) and
of hadronic energy (right scale, filled symbols) versus the core
distance for showers of truncated muon numbers as indicated. 
Threshold energy for hadrons is 50~GeV. The curves represent fits
of the NKG formula to the data at $r \ge 8$~m with a radius fixed
to $r_h$ = 10~m.}
\label{fig:had_density}
\end{figure}

Data on lateral distributions of hadrons studied over a large
range of distances to the shower core are very scarce in the
literature \cite{hasegawa65}.  The KASCADE calorimeter operated
jointly with the array detectors enables such investigations to
be performed with high quality.  Different from electrons and
muons, the reconstruction of individual energies of hadrons
enables to study in detail also the hadronic energy dependence of
lateral distributions as well as to compare lateral particle and
energy density distributions.  As an example,
Fig.~\ref{fig:had_density} presents hadron lateral distributions
for four $\Nmutr$ sizes corresponding approximately to the energy
interval from 1 to 10~PeV. The densities of hadrons and of
hadronic energy are given.  They extend up to distances of 90~m
from the shower core where the intensity has dropped by nearly
five orders of magnitude.  At the very centre, a saturation as
mentioned in section~\ref{sec:hadrec} can be noticed for the
hadron number.  Hence, in this range the hadronic energy is the
more reliable observable.

\begin{figure}
\centerline{\epsfig{file=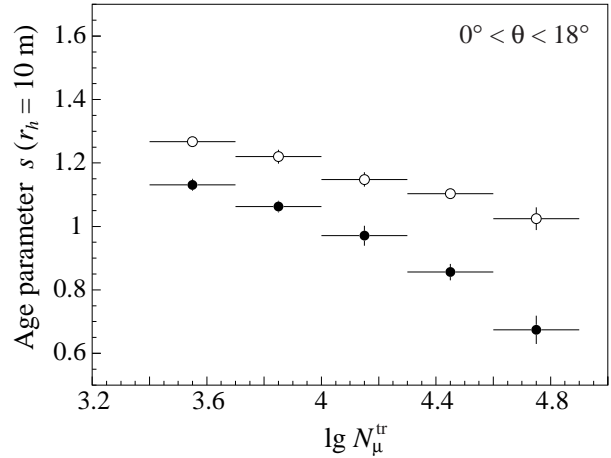,width=8cm}}
\caption{The age parameter $s$ for fixed radius $r_h$ = 10~m versus 
the muon shower size $\Nmutr$. Open symbols are for hadronic 
particle density and full symbols for hadronic energy density.} 
\label{fig:had_age}
\end{figure}

Several functions have been applied to fit the data points, among
others exponentials suggested by Kempa \cite{Kempa-1976}. 
However, by far the best fit was obtained when applying the NKG
formula represented by the curves shown in the graph.  Because of
the mentioned saturation effects close to the shower centre, the
actual fit is only applied to the data points within the range of
the full lines while the dashed curves are extrapolations to
smaller distances.  The distributions are much narrower than
those of the electrons and the scale radii determined by the fit
are about $r_{h} \simeq 10$~m.  Furthermore, a variation of the
lateral shape is observed which is similarly to that of the
electron LDF. When fixing $r_h$ = 10~m to determine the age
parameters $s$, we get the result presented in
Fig~\ref{fig:had_age}.  The age parameter yields values similar
to the electromagnetic and muonic component and decreases with
increasing shower size as expected.

The smaller scale radii and the observed variation with shower
size may be interpreted in the picture of high energy hadrons
passing through the atmosphere and generating essentially the
electromagnetic component.  Multiple scattering of electrons then
resembles the scattering character of hadrons with a mean
transverse momentum of 400~MeV/c almost irrespective of their
energy.  Hence, in a dimensional estimate of the hadronic lateral
scale radius we substitute in the formula of the Moli\`{e}re
radius, $\Rmol = X_{0} \cdot E_{s} / E_{c}$, the radiation length
$X_{0}$ by the hadronic interaction length, the scaling energy
$E_{s} =m_{e}c^{2} \sqrt{4\pi/\alpha} \simeq 21.2$~MeV by the
mean transverse momentum, and the critical energy $E_{c}$ (which
approximately coincides with the average energy of the electrons
at observation level) by the threshold energy of detected hadrons
and arrive at a radius $r_h \cong {\rm 1.2~km} \times {\rm
400~MeV/50~GeV} \cong 10$~m, such as is observed experimentally.

\begin{figure}
\epsfig{file=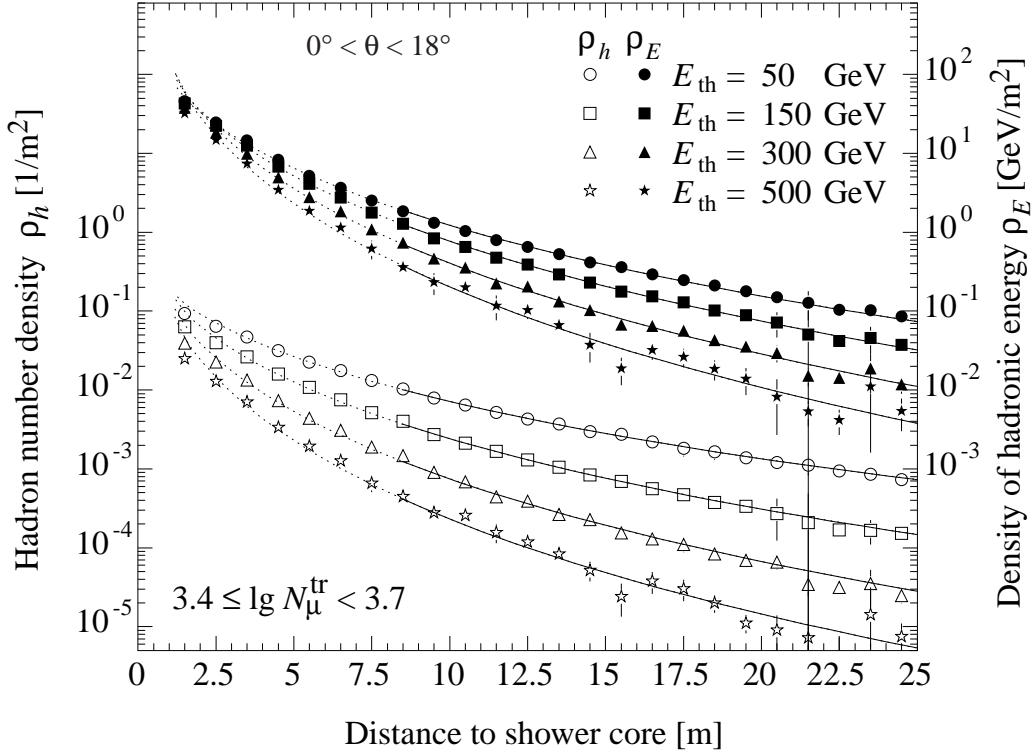,width=\textwidth}
\caption{Density of hadron number (left scale, open symbols) and
of hadronic energy (right scale, filled symbols) versus shower
core distance for various thresholds of hadron 
energy.
The curves represent fits of the 
data to the NKG function as in Fig.~\ref{fig:had_density}.}
\label{fig:had_innerdensity}
\end{figure}

\begin{figure}
\centerline{\epsfig{file=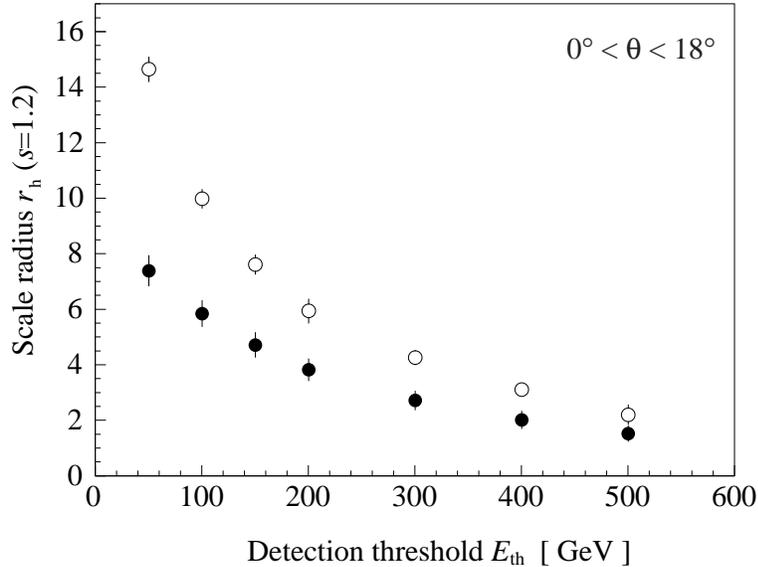,width=10cm}}
\caption{Hadronic scale radius $r_{h}$ as a function of 
detection threshold $E_{\rm th}$ for a fixed shower 
age of $s=1.2$ and $3.4 < \lg N_{\mu}^{\rm tr} < 3.7$. Filled symbols 
are for hadronic energy density and open symbols for hadronic 
particle density.}
\label{fig:had_variation}
\end{figure}

Figure~\ref{fig:had_innerdensity} provides a closer view to the
shower axis for different threshold energies.  Again, hadron
numbers and energy densities are given.  We observe that
energetic hadrons are concentrated very close to the centre.  The
number of TeV hadrons drops by an order of magnitude within the
first 3~m.  The energy density $\rho_{E}$ is well described by
the NKG formula down to small distances to the shower axis. 
Deviations at distances up to about 1~m are attributed to the
limited core position resolution of the detector.  The variation
of the hadronic scale radius, $r_{h}$, with the detection
threshold, $E_{\rm th}$, applied to the hadrons is displayed in
Fig.\,\ref{fig:had_variation}.  The data corroborate the expected
dependence of the shower width on energy threshold, $E_{\rm th}$,
as outlined above.

\section{Summary and Outlook}

Measurements of electron, muon, and hadron lateral distributions
as recorded by the KASCADE experiment have been presented for
radial distances of up to 200~m and for the energy range
$5\times10^{14} {\rm eV} < E < 10^{17} {\rm eV}$.  Detector
simulations were performed to account for effects like muon and
hadron contamination in signals of the $e/\gamma$-scintillators
and for punch-through of electrons and hadrons into the muon
detectors.
          
All types of lateral distributions are well described by
NKG-functions using different scale radii $r_{i}$ (with
$i=e,\mu,h$) for the different air shower components.
\begin{itemize}

\item A study of the electron LDFs shows that optimum
fits are not obtained for the canonical value of $\Rmol = 79$~m,
but for $r_{e} \simeq 20-30$~m, i.e.\ we observe a stronger
curvature in the experimental data than in the conventional NKG
function.  This imposes a systematic effect of up to 5\,\% in the
integrated number of shower electrons.  Due to the strong
correlation of $r_{e}$ and $s$, the preferred lower scale radius
is accompanied by a larger age parameter of $s \simeq 1.65$.  The
optimum set of parameters depends on shower size and zenith angle
and may be used to infer the mass of the primary particle. 
For practical reasons and because of limited statistics within
single events, information about the shape of the electron LDF is
usually extracted by fixing $r_{e}$ and fitting only the age
parameter $s$.  A problem specific to NKG functions in this
approach and with small scale radii $r_{e}$ is, that upwards
fluctuations of $s$ easily lead to ill defined shower sizes.  As
an alternative, we have demonstrated that a fixed age parameter
but variable scale radius provides an equally good fit to the
data.  The parameter $r_{e}$ then changes in a characteristic way
and also exhibits a distinct structure at shower sizes
corresponding to the knee position.

\item Within the fiducial area of KASCADE, muon LDFs are well
described by a NKG function, but with a scale radius of $r_{\mu}
= 420$~m.  Because of the limitation to 40~m $< r <$ 200~m, the
experiment is not very sensitive to the actual value chosen and
the data are also equally well described by a Greisen
parametrization.  Significant differences would only occur at
radial distances outside the acceptance of the experiment.  The
unknown flat shape of the muon LDF at large distances imposes
serious problems (even for much larger surface detector arrays)
when calculating the total number of muons within an air shower. 
Most importantly, $N_{\mu}$ is subject to systematic shifts and
increased fluctuations, thereby deteriorating the shower size and
primary energy resolution.  Thus, for classifying events, we have
introduced the truncated muon number, $N_{\mu}^{\rm tr}$,
obtained from integrating the LDF only within the experimental
acceptance of 40-200~m.  A rough scan of the low energy muon
spectrum has been performed by analysing LDFs at $E_{\mu} \ge
230$, 490, and 2400~MeV.
Similarly to electrons, a steepening of the
muon LDF is observed with increasing shower size and decreasing
zenith angle, as is expected for observations being increasingly
closer to the shower maximum.

\item Quite interestingly, also hadronic lateral energy density
and particle number distributions are well approximated by the
NKG form up to distances of at least 90~m.  The scale radius for
$E_{\rm th} \ge 50$~GeV is $r_{h} \simeq 10$~m and scales
roughly proportional to $E_{\rm th}^{-1}$, as expected by a simple
dimensional comparison of electromagnetic multiple scattering and
hadronic interactions.

\item The interrelation between the electromagnetic and hadronic
EAS component may explain the `unconventional' small preferred
scale radius of the electron LDF of 20-30~m as compared to the
classical value of $\Rmol \simeq 80$~m.  It should be kept in
mind that the classical Moli\`ere radius has been derived for
pure electromagnetic showers and for zero energy threshold only. 
However, extensive air showers are mostly initiated by primary
hadrons.  Therefore, the shower evolution is mostly driven by the
substantially narrower hadronic component, and the effective
lateral scale radius of observed electrons is expected to be
smaller than for $E_{\rm kin} \ge 0$ electrons in pure
$\gamma$-initiated showers.
\end{itemize}

%
The present paper is not focussed to detailed analyses in terms
of predictions of the EAS developments from Monte Carlo
simulations and to a comparison of different theoretical
high-energy interaction approaches like VENUS \cite{venus},
QGSJET \cite{qgsjet} and SIBYLL \cite{sibyll}.  These models,
continuously in the process of refinement, are generators
implemented into the Karlsruhe EAS Monte Carlo code CORSIKA
\cite{CORSIKA}.  However, the presented results provide a
coherent experimental basis for serious tests considering
simultaneously the three main EAS components, not only concerning
the interaction but also the particle propagation procedures.  It
may be noted that the muon lateral distributions are
experimentally given for three different energy detection
thresholds of the registered muons, thus implying also some
sensitivity to the low energy spectrum.  Most valuable for such
tests are observations based on the hadronic component.  An
example of first analyses in this scope were presented in
\cite{kascade-ww} and a remarkable agreement of lateral
distributions of hadrons for primary protons and Fe nuclei was
observed.  In particular, the absence of peculiar features, in
contrast to earlier observations by Danilova et al.\
\cite{Danilova-1985} and Arvela and Elo \cite{Arvela-1995} can be
stated, even at energies as high as 10 PeV. Such results support
the trust in a correct handling of the particle propagation and
of the development of the hadronic component at least for hadron
energies above 50 GeV. More detailed comparisons of lateral
distributions with CORSIKA simulations are under study and will
be subject of a forthcoming publication.

\section*{Acknowledgements}
\vspace*{-8mm}
The authors are indebted to the members of the engineering and 
technical staff of the KASCADE collaboration, who contributed with 
enthusiasm and engagement to the success of the experiment.

The support, based on common projects of Scientific-Technological
Cooperation Agreements (WTZ) and provided by the Ministery for
Research of the German Federal Government and International
Bureau Bonn, is gratefully acknowledged.  The collaboration has
been partly supported by grants of the Polish Committee for
Scientific Research, the Romanian Ministery of Research and
Technology, of the Armenian Government and an ISTC project
(A116).

\bibliographystyle{nim}
\bibliography{Lateral}

\end{document}